\documentclass[%
 reprint,
 amsmath,amssymb,
 aps,
prb,
]{revtex4-2}

\usepackage{graphicx}
\usepackage{dcolumn}
\usepackage{bm}
\usepackage{adjustbox}
\usepackage{hyperref}
\DeclareUnicodeCharacter{2212}{-}
\DeclareUnicodeCharacter{03B4}{$\delta$}

\begin{document}

\preprint{APS/123-QED}

\title{Anisotropic Nonsaturating Magnetoresistance Observed in HoMn$_6$Ge$_6$: A Kagome Dirac Semimetal}

\author{Achintya Low}

\author{Tushar Kanti Bhowmik}

\author{Susanta Ghosh}

\author{Setti Thirupathaiah}%
 \email{setti@bose.res.in}
\affiliation{%
 Department of Condensed Matter Physics and Material Sciences, S. N. Bose National Centre for Basic Sciences, Kolkata-700106
}%


\begin{abstract}
We report the magnetic and magnetotransport properties and electronic band structure of the kagome Dirac semimetal HoMn$_6$Ge$_6$. Temperature-dependent electrical resistivity demonstrates various magnetic-transition-driven anomalies. Notably, a crossover from negative to positive magnetoresistance (MR) is observed at around 150 K. While the linear nonsaturating positive MR in the low-temperature region is mainly driven by the linear Dirac-like band dispersions as predicted by the first-principles calculations, the negative MR observed in the high-temperature region is due to the spin-flop type magnetic transition. Consistent with anisotropic Fermi surface topology, we observe anisotropic magnetoresistance at low temperatures. A significant anomalous Hall effect has been noticed at high temperatures in addition to a switching of the dominant charge carrier from electron to hole at around 215 K.
\end{abstract}

\maketitle


\section{Introduction}

The combination of frustrated magnetism, strong electronic correlations, and quantum topology in the kogome systems immensely attracts researchers from the fundamental science and potential technological applications points of view~\cite{Brooks1991, Haldane1988,Chang2013,Keimer2017,Sachdev2018,Tokura2019,Canfield2019, Hirohata2020}. Particularly, the interplay among these peculiar physical properties in the kagome lattice features intriguing quantum phenomena, including the quantum spin liquid phase~\cite{Kitaev2006,Helton2007,Han2012,Kasahara2018}, flat bands~\cite{Sutherland1986,Leykam2018,Kang2019,Kang2020}, Dirac and Weyl fermions~\cite{Nakatsuji2015,Kuroda2017,Liu2018,Ye2018,Liu2019,Kang2019}, Chern quantum phase~\cite{Xu2015,Yin2020}, and Skyrmionic lattice~\cite{Hirschberger2019,Chakrabartty2022}. So far, the transition metal-based kagome systems have been extensively explored to find the  Weyl fermions in Mn$_3$Sn(Ge)~\cite{Kuroda2017,Chen2021}, giant anomalous Hall effect in Co$_3$Sn$_2$S$_2$~\cite{Liu2018}, Nernst effect in Fe$_3$Sn~\cite{Chen2022}, Skyrmion lattice in Fe$_3$Sn$_2$~\cite{Hou2017}. On the other hand, the recent discovery of a nearly ideal quantum limit Chern magnet with massive Dirac fermion in TbMn$_6$Sn$_6$ has ignited an intense search for rare-earth-based kagome systems~\cite{Yin2020}.

Following this discovery, several experimental reports appeared discussing the tuning of topological properties in RMn$_6$Sn$_6$ (R = Gd-Tm, Lu)~\cite{Ma2021}, large anomalous Hall effect in RMn$_6$Sn$_6$ (R = Tb, Dy, and Ho)~\cite{Gao2021}, Chern magnetic state in TbMn$_6$Sn$_6$~\cite{Yin2020}, Dirac-like band crossings in HoMn$_6$Sn$_6$~\cite{Kabir2022}, linear unsaturated magnetoresistance in RMn$_6$Sn$_6$ (R = Gd, Tm, Lu)~\cite{Ma2021}. Among these, HoMn$_6$Sn$_6$ is a ferrimagnetic metal in which the spins of the Ho-$4f$ sub-lattice align antiferromagnetically with the spins of the Mn-$3d$ sub-lattice. Thus, the interplay between the ferrimagnetism and the electronic band structure plays a vital role in shaping the linear non-saturated magnetoresistance of this system~\cite {Campbell2021,Ma2021,Kabir2022}. On the other hand, though HoMn$_6$Ge$_6$ shares a similar crystal structure to HoMn$_6$Sn$_6$, minimal information is available on the magnetotransport properties of HoMn$_6$Ge$_6$ discussing in detail the relation between magnetic and magnetotransport properties~\cite{Zhou2023}.

HoMn$_6$Ge$_6$ crystallizes into HfFe$_6$Ge$_6$-type hexagonal structure with a space group of $P6/mmm$, ordering antiferromagnetically at a N$\acute{e}$el temperature of 466 K~\cite{Venturini1992}. The magnetic moments of Mn and Ho are arranged in a skew-spiral ($\widetilde{SS}$) fashion, and the plane of magnetic moments (Ho and Mo) is angled at 60$^o$ with the $z$-axis in HoMn$_6$Ge$_6$~\cite{SchobingerPapamanteilos1995}. In contrast, in $R$Mn$_6$Sn$_6$ systems, the moments of $R$ and  Mn are aligned antiparallel to form a ferrimagnetic structure with an easy-axis of magnetization angled with the $z$-axis~\cite{Yin2020,Gao2021,Ma2021,Kabir2022}. The angle between the $z$-axis and the easy-axis of magnetization strongly depends on the rare-earth element~\cite{SchobingerPapamanteilos1995, Yin2020,Gao2021,Ma2021,Kabir2022, Lee2023}.   Further, as discussed in Ref.~\cite{SchobingerPapamanteilos1995}, the magnetic structure of these systems is very sensitive to the sample temperature. For instance, in the low-temperature region ($<$ 55 K), the Ho and Mn magnetic moments form a skew-spiral ($\widetilde{SS}$) structure, coupled antiferromagnetically, with a plane of moments making an angle $\theta_S$ with the $z$-axis, producing finite net magnetization along both the in-plane and out-of-plane directions. However, as the temperature is increased, the skew-spiral structure gets distorted, and the other magnetic transitions emerge at 220 and 260 K. Beyond 260 K, the $\widetilde{SS}$ structure completely gets destroyed, and the system enters a cycloid state beyond 300 K~\cite{SchobingerPapamanteilos1995}.

\begin{figure}[t]
\includegraphics[width=0.9\linewidth]{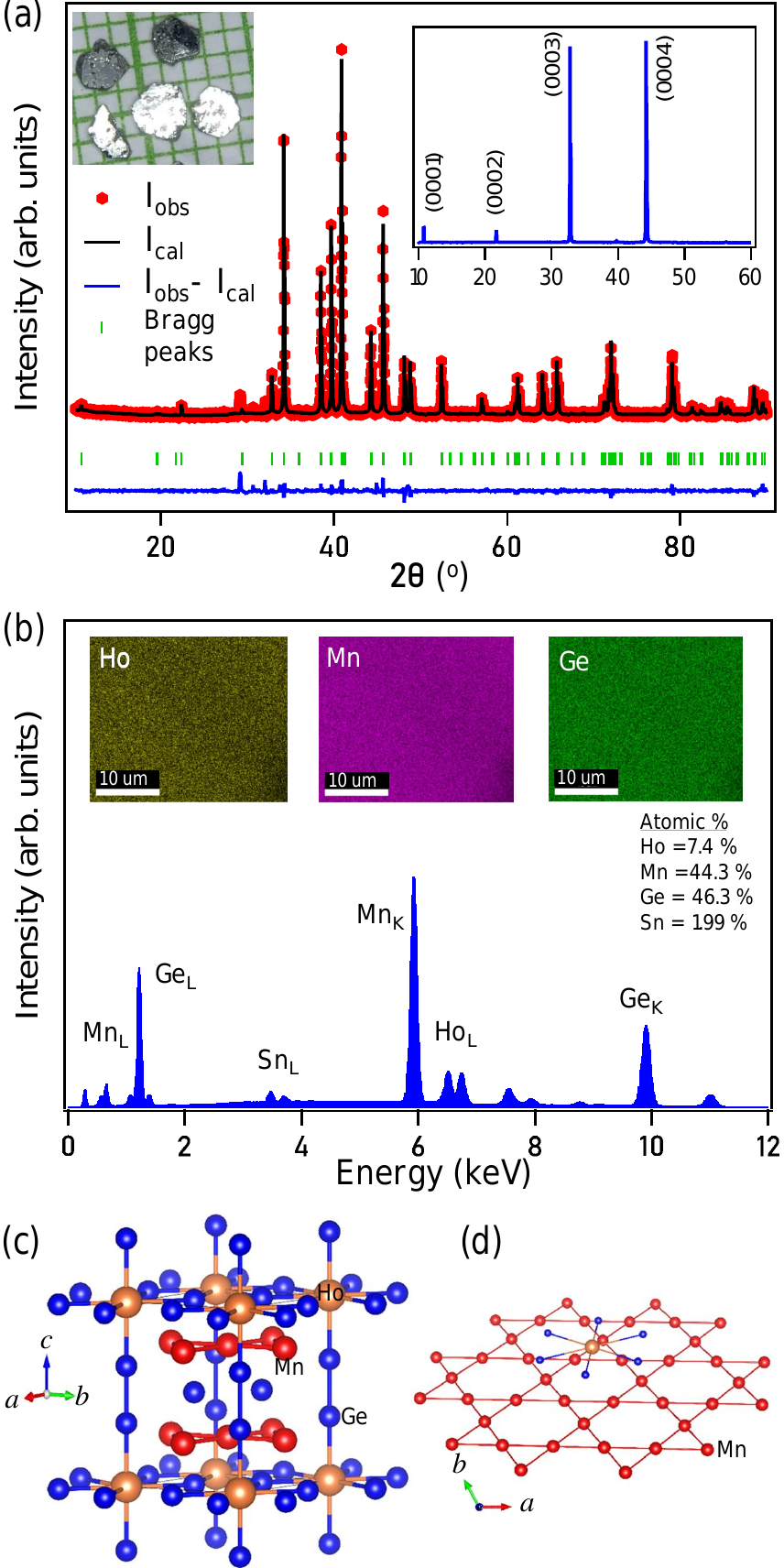}
\caption{\label{fig:Fig1} (a) The powder XRD pattern taken on the crushed single crystals of HoMn$_6$Ge$_6$ overlapped with Rietveld refinement. The left inset in (a) shows an optical image of as-grown single crystals. The right inset in (a) represents the XRD pattern corresponding to the (000l) Bragg plane of the HoMn$_6$Ge$_6$ single crystal. (b) EDXS data taken on HoMn$_6$Ge$_6$ single crystal. Insets in (b) show the elemental mapping of Ho, Mn, and Ge atoms.  (c) Schematic representation of the primitive unit cell of HoMn$_6$Ge$_6$. (d) Kagome lattice plane formed by the Mn atoms when projected onto the $ab$-plane.}
\end{figure}

In this work, high-quality single crystals of HoMn$_6$Ge$_6$ were grown using the Sn flux to study the magnetic and magnetotransport properties. Electrical resistivity demonstrates an overall metallic nature throughout the measured temperature range with a few magnetic transition-driven anomalies. A crossover from negative to positive magnetoresistance (MR) is observed at a critical temperature of 150 K. While the Dirac-like linear band dispersion mainly drives the linear nonsaturating positive MR in the low-temperature region, the negative MR observed in the higher temperature region is due to the spin-flop type magnetic transition. Most importantly, we identify a large anisotropy in the magnetoresistance due to the anisotropic Fermi surface present in this system. Further, we observe anomalous Hall effect in addition to a switching of dominant charge carriers from electrons to holes at around 215 K, ingoing from high to low temperatures.  Also, we performed density functional theory calculations on HoMn$_6$Ge$_6$ by considering the experimentally obtained skew-spiral magnetic structure to understand our magnetotransport data better~\cite{SchobingerPapamanteilos1995}.

\section{Experimental and First-principles Calculation Details}

Single crystals of HoMn$_6$Ge$_6$ were grown using Sn flux. Firstly, Holmium (Ho) chunk (Alfa Aesar, $99.9\%$), Manganese (Mn) powder (Alfa Aesar, $99.95\%$), Germanium (Ge) powder (Alfa Aesar, $99.99\%$), and Tin (Sn) shots (Alfa Aesar, $99.995\%$) were weighed in the ratio of $1: 6: 6: 24$. The mixture was kept in an alumina crucible and sealed the crucible together with the mixture inside an evacuated quartz ampoule. The mixture was heated up to 1000$^o$C at a rate of 50$^o$/hr and kept there for another 18 hrs before cooling the molten mixture down to 600$^o$C at a rate of 5$^o$C/hr. Finally, the molten is annealed for another two days at 600$^o$C. Then, the ampoule was quickly transferred to a centrifuge to separate the Sn flux from the samples. In this way, we obtained several hexagonal-shaped HoMn$_6$Ge$_6$ single crystals with a typical size of 2mm $\times$ 2mm $\times$ 0.4mm as shown in the inset of Fig.~\ref{fig:Fig1}(a). X-ray diffraction (XRD) was done on flat-shaped single crystals and as well on the powders of crushed crystals using Rigaku SmartLab with 9 kW Cu K$_\alpha$ x-ray source. The exact chemical composition of the as-grown single crystals was found using the energy dispersive x-ray spectroscopy (EDXS). The electrical resistivity and magnetotransport measurements were performed using the four-probe and Hall-probe connections. Electrical, magnetic, and magnetotransport measurements were performed in a 9 T physical properties measurement system (PPMS, DynaCool of Quantum Design) using the VSM and ETO options. To eliminate the longitudinal voltage contributions due to a possible misalignment of the Hall connections, the Hall resistivity was calculated using the relation $\frac{\rho_{H}(H)-\rho_{H}(-H)}{2}$.

To understand the magnetotransport properties, the electronic band structure of HoMn$_6$Ge$_6$ was calculated using the density-functional theory (DFT) within the Perdew-Burke-Ernzerhof-type generalized-gradient approximation (GGA)~\cite{Perdew1996} as implemented in the Quantum Espresso (QE) simulation package~\cite{Giannozzi2009}.  The electronic wavefunction is expanded using plane waves up to a cutoff energy of 100 Ry. Brillouin zone sampling is done over a 10$\times$10$\times$6 Monkhorst-Pack $k$-grid. The crystal structure was optimized through the variable-cell relaxation method as implemented in QE. Notably, the non-collinear skew-spiral magnetic structure was considered to determine the ground state electronic properties. The band structure was produced with and without considering the spin-orbit coupling (SOC).

\section{Results and Discussions}
Figure~\ref{fig:Fig1}(a) depicts the x-ray diffraction (XRD) pattern of the crushed HoMn$_6$Ge$_6$ single crystals overlapped with Rietveld refinement performed using the Fullprof software.  The Rietveld refinement confirms that HoMn$_6$Ge$_6$ crystallizes into the hexagonal kagome HfFe$_6$Ge$_6$-type crystal structure with a space group of $P6/mmm$ (No. 191). The refined lattice parameters $a=b=5.2415(5)~\AA$ and $c=8.1831(4)~\AA$ confirm that the studied sample of HoMn$_6$Ge$_6$ is almost equivalent to the reported ones within the error-bars~\cite{SchobingerPapamanteilos1995, Zhou2023}. Also, see Table I in the Supplemental Material~\cite{Supple} for a comparison of the structural parameters between our sample and the reported similar compounds in Refs.~\cite{SchobingerPapamanteilos1995, Zhou2023}. The right-inset in Fig.~\ref{fig:Fig1}(a) shows the XRD pattern taken on the flat surface of the single crystal, confirming that the crystal growth plane is parallel to (0001)-plane direction. Fig.~\ref{fig:Fig1}(b) depicts the EDXS data conforming the exact chemical composition of the sample Ho$_{0.96}$Mn$_{5.74}$Ge$_{6}$. For convenience, hereafter, we represent it by the nominal composition of HoMn$_6$Ge$_6$. Since the crystals were grown out of Sn flux, the crystals were found with a maximum 2\% of Sn impurity.   The elemental mappings done for Ho, Mn, and Ge, as shown in the top insets of Fig.~\ref{fig:Fig1}(b), imply a good homogeneity of the studied single crystals.

\begin{figure}
\includegraphics[width=\linewidth]{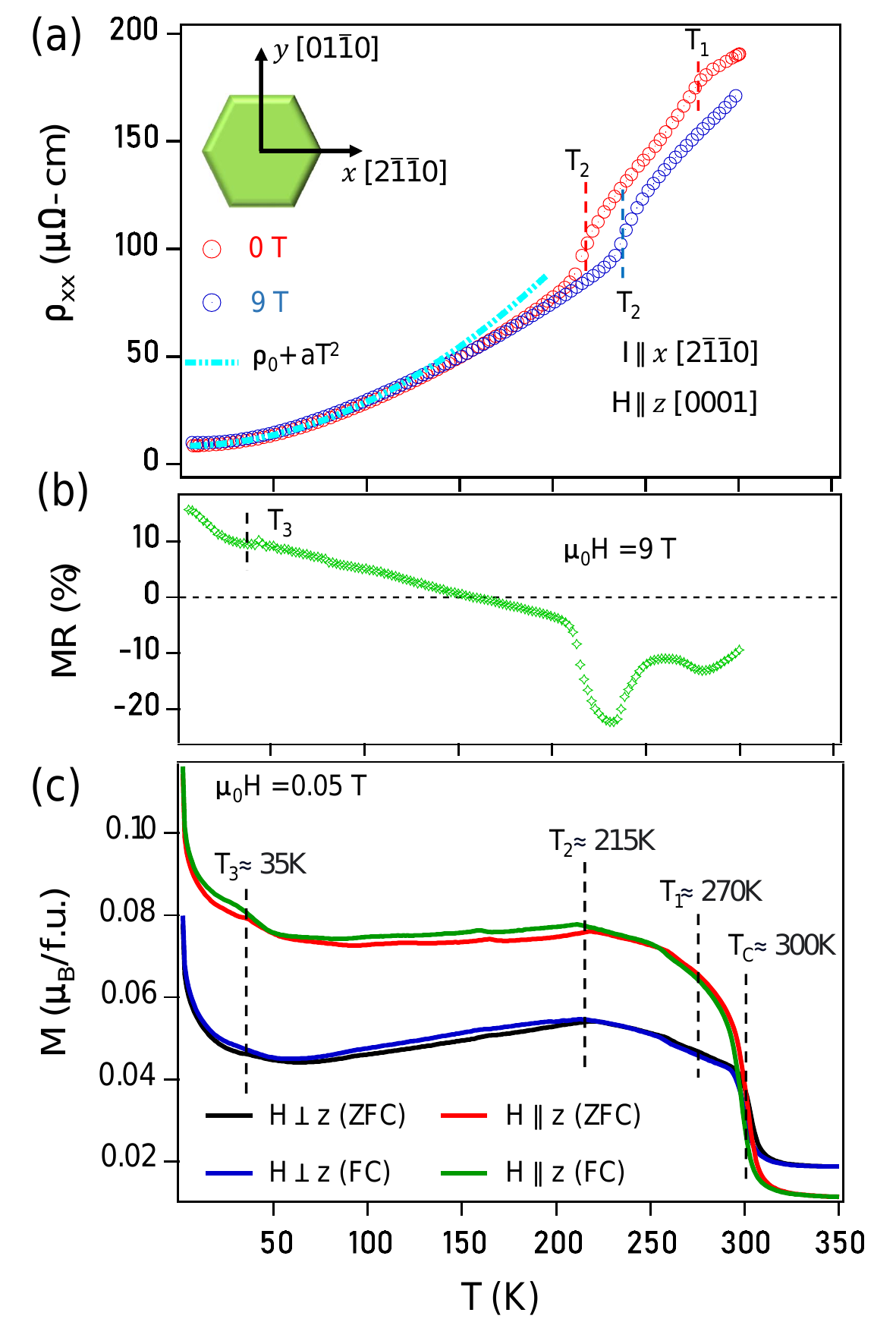}
\caption{\label{fig:RT_MT} (a) Longitudinal resistivity $\rho_{xx}$ plotted as a function of temperature with current applied along $x$ $[2\bar{1}\bar{1}0]$-axis for both 0 T (red curve) and 9 T (blue curve) fields applied along $z$ $[0001]$-axis of the crystal. (b) Magnetoresistance MR(\%)$=\frac{\rho_{xx}(9T)-\rho_{xx}(0T)}{\rho_{xx}(0T)}\times 100(\%)$ plotted as a function of temperature. (c) Magnetization plotted as a function of temperature [$M(T)$] for both $H\parallel z$ and $H\perp z$ directions measured under field-cooled and zero-field-cooled modes with  $\mu_0H=0.05$ T. }
\end{figure}

Longitudinal electrical resistivity [$\rho_{xx}(T)$] is plotted as a function of temperature, as shown in Fig.~\ref{fig:RT_MT}(a), measured without magnetic field (red-colored data) and with $\mu_0H=9$ T of magnetic field (blue-colored data) applied parallel to the $z$-axis [0001]. The low-temperature resistivity ($<$ 100 K) can be reasonably fit with the quadratic equation of the Landau Fermi-liquid theory, $\rho_{xx}(T)=\rho_0+\alpha T^2$. Here,  $\rho_0=8.42 $ $\mu\Omega cm$ is the residual resistivity due to impurity scattering and  $\alpha=2\times10^{-9}$ $\mu\Omega cm/K^2$ is the electron-electron scattering coefficient. Further, in the resistivity data ($\mu_0H=0$ T) we observe $kinks$ at $T_1\approx$ 270 K and $T_2\approx$ 215. The transition temperatures $T_1$ and $T_2$ shift towards higher temperatures under the external magnetic fields. For instance, at a field of 9 T, the transition temperature $T_2$ shifted to 236 K, while we cannot identify $T_1$ under the 9 T field as the measurements were done only up to 300 K. This observation indicates the magnetic origin of the $kinks$. Further, the observation of $T_1$ and $T_2$ transition temperatures under zero field  is consistent with a previous report on HoMn$_6$Ge$_6$ in which the temperature range between $T_1$ and $T_2$ is referred by a mixed magnetic state ($\widetilde{SS}$+AFM)~\cite{Zhou2023}.   Fig.~\ref{fig:RT_MT}(b) depicts the magnetoresistance percentage [$MR\%=(\rho_{xx}(T,H)-\rho_{xx}(T,0))/\rho_{xx}(T,0)$] plotted as a function of temperature from which another transition is identified at $T_3\approx$ 35 K, though it is not clearly visible from the $\rho_{xx}(T)$ data [see Fig.~\ref{fig:RT_MT}(a)]. Overall, the longitudinal electrical resistivity shows the metallic nature throughout the measured temperature range with a residual resistivity ratio (RRR) of $\rho_{xx}$(300K)/$\rho_{xx}$(2K) $\approx$ 20.

\begin{figure*}
\includegraphics[width=\linewidth]{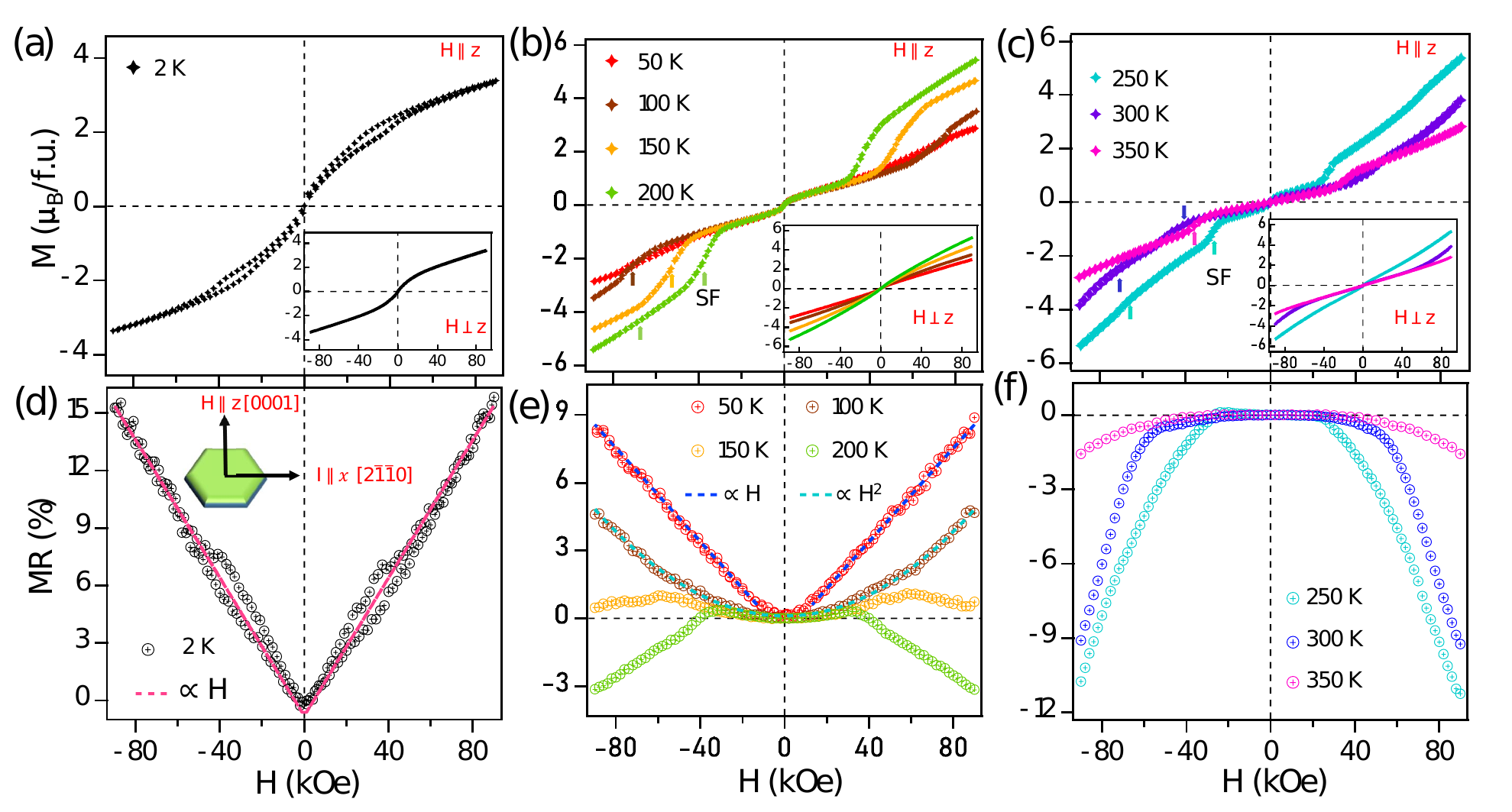}
\caption{\label{fig:MR_MH} (a),(b), and (c) Isothermal magnetization $M(H)$ measured at different temperatures for the fields applied parallel to the $z$-axis ($H \parallel z$). Insets in (a)-(c) show same for $H \perp z$. (d),(e), and (f) Field dependent magnetoresistance [MR(\%)$=\frac{\rho_{xx}(H)-\rho_{xx}(0)}{\rho_{xx}(0)}\times 100(\%)$] measured for different temperatures.}
\end{figure*}

Fig.~\ref{fig:RT_MT}(c) presents the magnetization plotted as a function of temperature [$M(T)$] under the magnetic field $\mu_0H=$ 0.05 T applied parallel ($H\parallel z$) and perpendicular ($H\perp z$) to the $z$-axis, measured both in the field-cooled (FC) and zero-field-cooled (ZFC) modes. Clear overlapping between FC and ZFC data suggests that the magnetic moments are thermally reversible in the system throughout the measured temperature range of 2 - 350 K.  On the other hand, previous neutron diffraction study on similar system revealed a complex magnetic structure with different magnetic states at different temperatures~\cite{SchobingerPapamanteilos1995}. For instance, in the low-temperature region ($<$ 55 K), the Ho and Mn magnetic moments form a skew spiral ($\widetilde{SS}$) structure, coupled antiferromagnetically,  with a plane of moments making an angle $\theta_S$ with the $z$-axis [see Fig.~\ref{fig:DFT}(b)], producing finite net magnetization along both the in-plane and out-of-plane directions. Between 55 and 220 K, both wave vector ($q$) and phase angle ($\phi_S$) increase monotonically with temperature. In the 220-260 K region, the $\widetilde{SS}$ structure gets distorted, accompanied by the Mn spin-reorientation, and thus eventually leading to a weak ferromagnetism at 300 K. Beyond 300 K the skew-spiral structure becomes a cycloid. In line with previous studies~\cite{SchobingerPapamanteilos1995,Venturini2001,Mar2004, Zhou2023} we also observe magnetic transitions in the magnetization data at $T_1\approx$ 270 K, $T_2\approx$ 215 K,  and at $T_3\approx$ 35 K. Further, a weak ferromagnetic-like magnetization jump is noticed at $T_{C}\approx$ 300 K. As can be seen from Fig.~\ref{fig:RT_MT}(c), the out-of-plane magnetization ($H\parallel z$) is significantly higher than the in-plane magnetization ($H\perp z$) below 300 K, while it is vice versa above 300 K. This observation hints at the fluctuating magnetic moments across the $T_C$.  Also, note that a slight variation in the transition temperatures ($T_1$, $T_2$ and $T_3$ ) is noticed between our data and Refs.~\cite{SchobingerPapamanteilos1995, Zhou2023}, possibly due to different sample preparation methods. Specifically, the $M (T)$ of Ref.~\cite{Zhou2023} shows a cusp at around 25 K and a rapid increase in $M(T)$ below 77 K,  while no cusp is found but a rapid increase in $M(T)$ is found below 55 K in Ref.~\cite{SchobingerPapamanteilos1995}. On the other hand, from our data we observe a cusp at 33 K and a rapid increase in $M(T)$ below 50 K [see Fig.~\ref{fig:RT_MT}(c)]. Similarly, the mixed magnetic state of AFM+$\widetilde{SS}$ is reported within the temperature range of 220-260 K in Ref.~\cite{SchobingerPapamanteilos1995}, 215-245 K in Ref.~\cite{Zhou2023}, while from our data we find it in a broader temperature range of 215-270 K..

Next, Figs.~\ref{fig:MR_MH}(a), ~\ref{fig:MR_MH}(b), and ~\ref{fig:MR_MH}(c) depict magnetization isotherms [$M(H)$] measured at different temperatures for  $H\parallel z$ and $H\perp z$. We observe weak ferromagnetic-like $M(H)$ data when measured at 2 K for both $H\parallel z$ and $H\perp z$ at lower field regions without magnetization saturation. Further field-induced hysteresis is visible for $H\parallel z$ between 2 and 4 T but not for $H\perp z$, possibly due to the out-of-plane spin canting within this field range. This type of metamagnetic state is not observed at higher temperatures. Instead, an out-of-plane spin-flop transition has been found at a critical field ($H_{SF}$) that is temperature dependent, due to which a sudden increase in magnetization was noticed [see Figs.~\ref{fig:MR_MH}(b) and(c)]. However, the spin-flop transition is absent from the in-plane magnetization ($H\perp z$) at any measured temperature. A previous report also demonstrates similar spin-flop transitions in  HoMn$_6$Ge$_6$~\cite{Zhou2023}.

Isothermal magnetoresistance (MR) at different sample temperatures and with the fields applied parallel to the $z$-axis are shown in Figs.~\ref{fig:MR_MH}(d), ~\ref{fig:MR_MH}(e), and ~\ref{fig:MR_MH}(f). Fig.~\ref{fig:MR_MH}(d) demonstrates the MR measured at 2 K, where the MR is linearly dependent (red-dashed line) on the applied field in addition to the hysteresis between the fields 2 and 4 T that originated from the field-induced metamagnetic state [Fig.~\ref{fig:MR_MH}(a)]. From Fig.~\ref{fig:MR_MH}(e), we can see that the linear dependency of MR is intact up to 50 K. However, at 100 K, we observe a classical parabolic magnetoresistance. Though this parabolic nature is sustained at higher temperatures (150 and 200 K) as well, beyond the spin-flop ($H_{SF}$) transition, the MR starts to decrease with increasing the field for a given temperature. Eventually, beyond the $H_{SF}$ the MR becomes negative above  150 K, consistent with the MR(T) data shown in Fig.~\ref{fig:RT_MT}(b) in which we can notice negative MR above 150 K when measured with 9 T. However, the MR becomes nearly field-independent as the temperature reaches 350 K.

Overall, HoMn$_6$Ge$_6$ shows a complex temperature dependent MR. Importantly, up to 50 K it shows a positive nonsaturating MR that is linearly depending on the applied field. There exist several mechanisms explaining the linear MR, such as (i) the presence of linear dispersive Dirac-like bands with very low effective mass near the Fermi level in the case of topological materials~\cite{Abrikosov1998,Abrikosov2000}, (ii) quasi-random resistor network model in the case of metal–semiconductor composites~\cite{Xu2008,Ramakrishnan2017}, and (iii) carrier density fluctuations due to irregular current paths from the inhomogeneous or grain boundaries in the case of disordered systems~\cite{Xu1997,Parish2003}. We can safely rule out the (ii) mechanism as our system is not a metal-semiconductor composite. Also, since our studied system is in the single crystalline form with no significant grain boundaries and with good homogeneity, we can rule out the mechanism (iii) also as the origin of linear MR. Previous reports on its sister compounds RMn$_6$Sn$_6$ suggested that the linear MR could originate from the Dirac-like linear bands~\cite{Ma2021}. In addition, previous band structure calculations on these systems showed several Dirac-like linear band dispersions crossing the Fermi level~\cite{Lee2023,Dhakal2021}. To confirm that the linear MR originates from the linear Dirac-like topological band structure, we performed density functional theory (DFT) calculation, as discussed later explicitly.

\begin{figure}
\includegraphics[width=\linewidth]{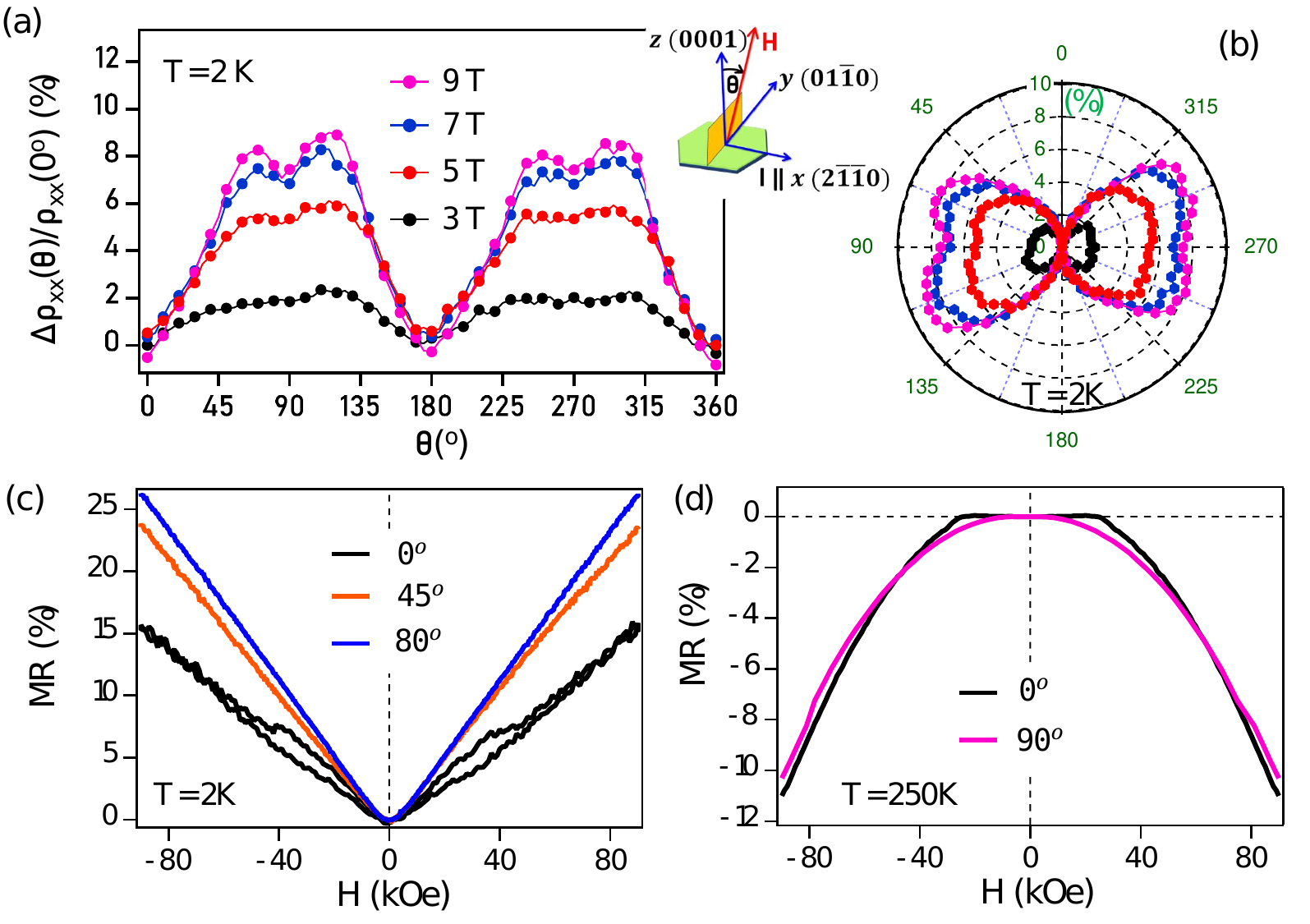}
\caption{\label{fig:RH_Th} (a) Angle dependent magnetoresistance [ADMR(\%) $=\frac{\rho_{xx}(\theta)-\rho_{xx}(0^o)}{\rho_{xx}(0^o)}\times 100 (\%)$] measured for various applied fields at 2 K. Inset in (a) is a schematic showing the ADMR measuring geometry.  Here $\theta$ is the angle between $z$-axis and the applied field direction. (b) is same as (a) but plotted in polar coordinates for a better visualization. In (b) radius is the amplitude of ADMR(\%). (c)  Field dependent magnetoresistance (MR) measured at 2 K for different $\theta$ values. (d) is same as (c) but measured at 250 K of the sample temperature.}
\end{figure}

Further, we measured the angle-dependent magnetoresistance (ADMR) with the current and field directions always kept perpendicular to each other such that only the field direction changes with the crystal axes as shown in the schematic of Fig.~\ref{fig:RH_Th}(a).   The angle-dependent magnetoresistance is calculated as $\frac{\rho_{xx}(\theta)-\rho_{xx}(0^o)}{\rho_{xx}(0^o)}$(\%) while the applied magnetic field is kept constant. As seen from Fig.~\ref{fig:RH_Th}(a), at 2 K, the MR is positive for all the applied magnetic fields but very sensitive to the field angle, suggesting a strong anisotropic MR in this system. Specifically, MR is minimum at the angles of 0$^\circ$, 180$^\circ$, and 360$^\circ$ and it is maximum at the angles 60$^\circ$, 120$^\circ$, 240$^\circ$, and 300$^\circ$. Also, the saddle points in the ADMR are observed at 90$^\circ$ and 270$^\circ$ for all the applied fields. For a better visualization, we plotted the ADMR in polar coordinates as shown in Fig.~\ref{fig:RH_Th}(b), from which we can clearly observe a butterfly pattern for the out-of-plane MR at 2 K. It is worth to mention here that the ADMR measured at 250 K is negligibly small (not shown) compared to the ADMR taken at 2 K.  Next, Fig.~\ref{fig:RH_Th}(c) depicts the field-dependent MR plotted for different field angles (0$^\circ$, 45$^\circ$, and 80$^\circ$) measured at 2 K, from which we observe linear MR at all field angles. Moreover, we see that the MR reaches almost 25\% for $\theta=80^{\circ}$, and the hysteresis, which was present for $\theta=0^{\circ}$, has vanished at the other two angles. Fig.~\ref{fig:RH_Th}(d) depicts filed dependent MR taken at $0^\circ$ and $90^\circ$ filed angles, from which we observe the absence of spin-flop transition effect on the MR when measured at $90^\circ$, consistent with $M(H)$ for $H\perp z$.

\begin{figure}[b]
\includegraphics[width=\linewidth]{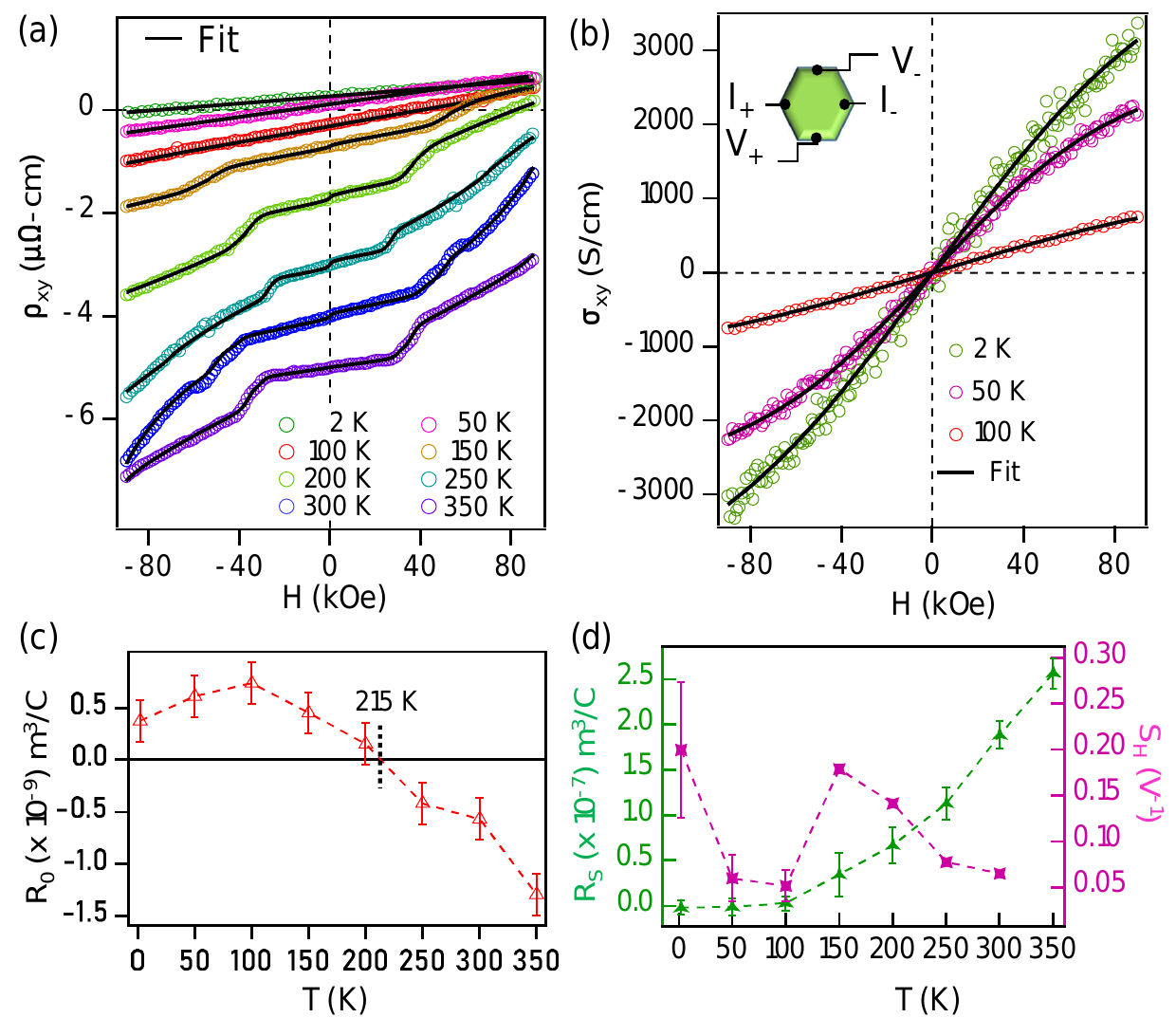}
\caption{\label{fig:RH_H} (a) Field dependent Hall resistivity $\rho_{xy}$ measured for different sample temperatures overlapped with the fits (black curves). (b) Field dependent Hall conductivity $\sigma_{xy}$ plotted for 2, 50, and 100 K of sample temperatures.  Inset in (b) schematically shows the Hall effect measuring geometry. (c) Normal Hall coefficient ($R_0$) plotted as a function of temperature. (d) Anomalous Hall coefficient ($R_S$) [left axis] and anomalous Hall scaling factor ($S_H$) [right axis] plotted as a function of temperature.}
\end{figure}

\begin{figure*}
\includegraphics[width=0.9\linewidth]{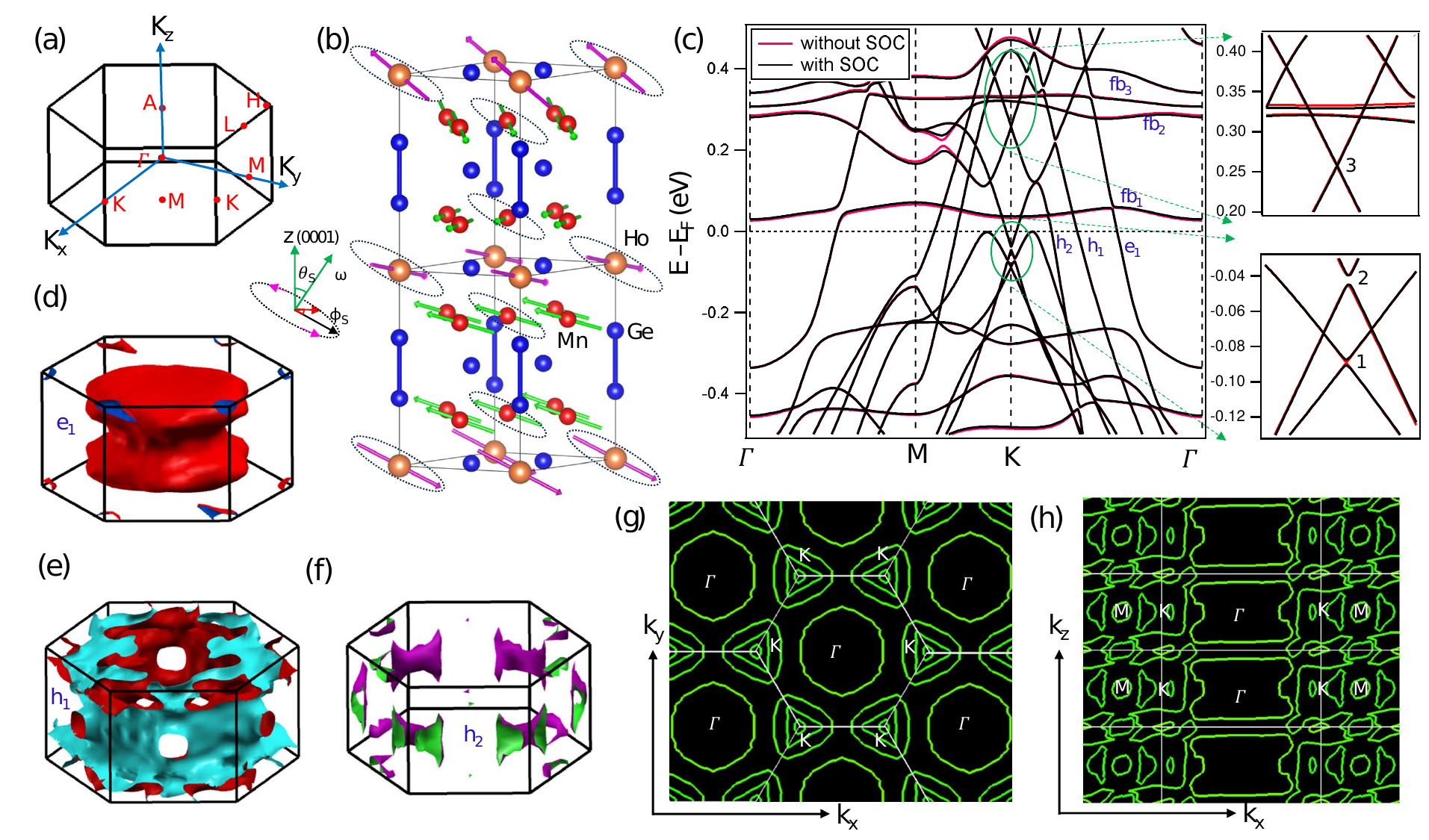}
\caption{\label{fig:DFT} (a) Hexagonal Brillouin zone showing the positions of various high symmetry points.  (b) Low temperature skew-spiral ($\widetilde{SS}$) magnetic structure of HoMn$_6$Ge$_6$~\citep{SchobingerPapamanteilos1995}. Here,  $\theta_s=60^o$ and $\phi_s=71.2^o$. (c) Electronic band structure of HoMn$_6$Ge$_6$ calculated for the magnetic configuration shown in (b) with (black-colored) and without (red-colored) including the spin-orbit coupling (SOC). The right-side panels of (c) show the zoomed-in band structure to clearly see various Dirac points. (d), (e), and (f) show the calculated Fermi surface maps in three-dimensional. (g) and (h) show the constant energy contours taken at the Fermi level in the $k_x-k_y$ and $k_x-k_z$ momentum planes, respectively.}
\end{figure*}

Next, Fig.~\ref{fig:RH_H}(a) demonstrate field-dependent Hall resistivity [$\rho_{xy}(H)$] measured at different sample temperatures. We see a deviation in the Hall resistivity, leading to an anomalous Hall effect,  at the spin-flop transition temperature. Usually, in ferromagnets, the Hall resistivity can be expressed by $\rho_H=R_0\mu_0H+R_S\mu_0M$~\cite{Nagaosa2010}, where $R_0$ is the ordinary Hall coefficient and $R_S$ is the anomalous Hall coefficient. Although the anomalous Hall effect (AHE) mainly appears in ferromagnets, recently it has been widely observed in noncollinear antiferromagnets as well~\cite{Nakatsuji2015,Luo2015,Suzuki2016,Niu2017}. From our Hall resistivity data, we can see that the anomalous Hall effect is induced by a spin-flop transition that resembles the $M(H)$ data shown in  Figs.~\ref{fig:MR_MH}(a)-(c). Moreover, the above formula can be rewritten as $\frac{\rho_H}{\mu_0H}=R_0+R_S\frac{M}{H}$ which imitates a linear equation and intercept on the $y$-axis gives $R_0$ and slope gives the anomalous Hall coefficient $R_S$. Using this formalism, we have reasonably fitted the Hall resistivity as depicted in Fig.~\ref{fig:RH_H}(a). The anomalous scaling factor ($S_H$) is calculated using the formula $S_H=\mu_0R_S/\rho_{xx}^2$, where $\rho_{xx}$ is the longitudinal resistivity. Fig.~\ref{fig:RH_H}(c) depicts the normal Hall coefficient ($R_0$) plotted as a function of temperature. From Fig.~\ref{fig:RH_H}(c), we further notice $R_0$ changing from positive to negative at around 215 K, which indicates hole-type (electron-type) carrier dominance below (above) 215 K.

As discussed above, a magnetic transition exists at around 215 K ($T_2$). It is possible that the change in magnetic structure influences the electronic structure near the Fermi level and thus the switching of charge carrier type at  215 K. This observation suggests a strong correlation between the magnetic and electronic structures in these systems. Moreover, the value of $R_S$ decreases rapidly with decreasing temperature and becomes negligibly small below 100 K. Whereas the $S_H$ varies between 0.05 and 0.2 like in a typical ferromagnetic metal~\cite{Nagaosa2010,Wang2016}. Further from the relation $\mu_0R_S=S_H\rho_{xx}^2$, it is clear that for very low resistivity $\rho_{xx}$ values the contribution from AHE ($R_S$) is negligible so that only the ordinary Hall effect dominates at low temperatures. In Fig.~\ref{fig:RH_H}(b), we have plotted the Hall conductivity as a function of temperature calculated using the formula, $\sigma_{xy}=-\frac{\rho_{xy}}{{\rho_{xy}^2}+{\rho_{xx}^2}}$. In Fig.~\ref{fig:RH_H}(b), we have fitted the Hall conductivity curves using the single band model, $\sigma_{xy}=[\frac{n_h{\mu_h}^2}{1+(\mu_hB)^2}]eB$ as at low temperatures the hole carriers dominate the total Hall transport. Here, $n_h$ is hole carrier density, and $\mu_h$ is hole mobility. From the fits, we derived the hole carrier concentration of $n_h$ is $1.06 \times 10^{20}$ cm$^{-3}$ and hole mobility of $\mu=0.049$ m$^2$/Vs at 2 K. These values are consistent with a previous report on these type of systems~\cite{Ma2021}.

To uncover the mechanism of unsaturated linear MR, we performed density functional theory calculations as illustrated in Fig.~\ref{fig:DFT}. For the DFT calculations, we considered the magnetic structure as shown in Fig.~\ref{fig:DFT}(b), depicted by the neutron diffraction study on this system~\cite{SchobingerPapamanteilos1995}. Fig.~\ref{fig:DFT}(c) displays the electronic band dispersion along the high-symmetry $k$-path of $\Gamma-\mathrm{M-K}-\Gamma$ calculated without and with considering the spin-orbit coupling (SOC). We notice several linear band crossings (Dirac-like) from the electronic band structure at the $K$ point near the Fermi level. One Dirac point (DP1) is observed at a binding energy of 90 meV below the Fermi level ($E_F$) that is gapless without SOC, but under SOC, the Dirac point is gapped by lifting the degeneracy. DP2 is found at a binding energy of 40 meV below the $E_F$, which is always gaped. DP3 is found at a binding energy of 250 meV above $E_F$, which is gapless for both with SOC and without SOC. Our calculations on HoMn$_6$Ge$_6$ align with previous calculations performed on HoMn$_6$Sn$_6$~\cite{Kabir2022, Lee2023}. Particularly, the calculations on  HoMn$_6$Sn$_6$ demonstrated that gap size at the Dirac points depends explicitly on the magnetic spin orientation with respect to the crystal axis~\cite{Kabir2022, Lee2023}.  This means, larger (smaller) gaps were predicted for the fields applied along the out-of-plane (in-plane) direction. Since our calculations were performed with an angle of $\theta_S=60^{\circ}$ between the $z$-axis and plane of magnetic moments, which is neither out-of-plane nor in-plane, we observe both gapped and gapless Dirac points. This observation hints at the possibility of a critical angle of magnetic moments at which the Dirac states are gapless and robust under SOC.

Apart from the linear band dispersions at the $K$ point, we also observe two hole-like band dispersions ($h_1$ and $h_2$) crossing the Fermi level around $K$ point and an electron-like ($e_1$) band dispersion crossing the Fermi level around the $\Gamma$ point. No bands are found at the M-point crossing the Fermi level. In addition, we found several flat bands ($fb_1$, $fb_2$, and $fb_3$) that are dispersionless throughout the Brillouin zone. Figs.~\ref{fig:DFT}(d)-(f) show the three-dimensional view of the Fermi surface maps, which are mainly contributed by three types of Fermi pockets; one of them is the electron-like ($e_1$) pocket with an almost cylindrical shape as shown in Fig.~\ref{fig:DFT}(d), the second one is the hole-like ($h_1$) pocket, which is close to one-third of a cylinder shared by each corner of the hexagon, as shown in Fig.~\ref{fig:DFT}(e), and the last one is the hip-roof-shaped hole-like ($h_2$) pocket at the $K$ point. With the help of these three types of Fermi pockets, we estimated the hole carrier density $n_h$=$5.3\times 10^{20}$/cm$^3$ and electron carrier density $n_e$=$3.6\times 10^{20}$/cm$^3$ to find the net hole carrier density of $1.7\times 10^{20}$/cm$^3$ using Luttinger's theorem~\cite{Luttinger1960} [see Supplemental Material~\cite{Supple}]. Interestingly, this value is close to the experimentally calculated net carrier density ($1.06 \times 10^{20}$ cm$^{-3}$) from the Hall data at 2 K. The estimated hole and electron carrier densities are nearly equal, suggesting a possible charge compensation in this system~\cite{Ali2014, Thirupathaiah2017}. Further, the estimated net carrier density of $1.06 \times 10^{20}$ cm$^{-3}$ suggests HoMn$_6$Ge$_6$ to be a semimetal~\cite{Qi2016, Thirupathaiah2017a}. Figs.~\ref{fig:DFT}(g) and ~\ref{fig:DFT}(h) illustrate the Fermi surfaces projected onto the $k_x-k_y$ and $k_x-k_z$ planes, clearly showing the electron-like Fermi pocket at the $\Gamma$ and hole-like Fermi pockets at the $K$ point.

As for the angle-dependent magnetoresistance (ADMR) measurements,  the magnetic field is always perpendicular to the current direction, which does not change the Lorentz force acting on the charge carriers. The spin-charge scattering can be neglected at low temperatures as the mean free path of the charge carriers is much higher in this system at 2 K ($\approx$ 0.67 $\mu m$) compared to the distance of $Mn-Mn$ (2.62 \AA) or $Ho-Ho$ (5.24 \AA)  magnetic moments. This leads us to conclude that the anisotropic ADMR has an electronic band structure origin rather than a magnetism origin. The Fermi velocity can be calculated as $v_k=\frac{1}{\hbar}\nabla_k\epsilon_k$, which depends on the local curvature of the Fermi surface cross-section. The smaller the orbit of the Fermi surface cross-section, the greater the local curvature, leading to a higher Fermi velocity. Hence, a high microscopic Lorentz force would act on the charge carriers~\cite{Collaudin2015,Wang2019,Zhang2019}. Our calculations indicate that the electron pocket $e_1$ and hole pocket $h_1$ exhibit nearly cylindrical shapes, with their diameters and heights being quite similar in value, making them nearly isotropic. Only the $h_2$ hole pocket has an anisotropic shape (hip-roof-shaped), which could be the possible origin of this high anisotropic ADMR. Because when the field is applied along the $z$-axis, the charge carriers are subjected to orbit on the $k_x-k_y$ plane. Similarly, for the fields applied along the $y$-axis the carriers would orbit on the $k_x-k_z$ plane. As observed in Fig.~\ref{fig:DFT}(g), the Fermi sheets at $K$ point are nearly circular for $H\parallel z$ [see Fig.~\ref{fig:RH_Th}(a) for $\theta=0^o$]. Only the out-of-plane Fermi sheets ($k_x-k_z$) have substantial curvature [see Fig.~\ref{fig:DFT}(h)] which are detected experimentally for $H\parallel y$ [see Fig.~\ref{fig:RH_Th}(a) for $\theta=90^o$]. Therefore, we think the asymmetric MR originated from the asymmetric out-of-plane Fermi pockets. Further, as evidenced by our DFT band structure calculations, the linear nonsaturating MR is mainly contributed by the linear Dirac-like bands near the Fermi level.

\section{Summary}

In summary, using the Sn flux, we have grown high-quality single crystals of HoMn$_6$Ge$_6$. Electrical resistivity, with a few magnetic transition-driven anomalies, demonstrates an overall metallic nature throughout the measured temperature range. A crossover from negative to positive magnetoresistance (MR) is observed at a critical temperature of 150 K. While the Dirac-like linear band dispersions mainly drive the linear nonsaturating positive MR exits in the low-temperature region, the negative MR observed in the higher temperature region is due to the spin-flop type magnetic transition. We found an anomalous Hall effect in addition to a dominant charge carrier switching across 215 K. We performed electronic band structure calculations on HoMn$_6$Ge$_6$ by considering the skew-spiral magnetic structure of the system to realize large anisotropy in the out-of-plane Fermi sheets. We suggest that the large anisotropic out-of-plane magnetoresistance observed in HoMn$_6$Ge$_6$ originated from the anisotropic out-of-plane Fermi surfaces. The band structure calculations predict several Dirac-type band crossings at the $K$ point near the Fermi level. 

\begin{acknowledgments}
This research has used the Technical Research Centre (TRC) Instrument Facilities of S. N. Bose National Centre for Basic Sciences, established under the TRC project of the Department of Science and Technology, Govt. of India.
\end{acknowledgments}

\nocite{*}

\bibliography{HMGref.bib}

\end{document}